\documentclass[10pt]{article}
\begin{document}
\title{Complete positivity for time-dependent qubit master equations}
\author{Michael J. W. Hall\\
Theoretical Physics, IAS, \\ Australian National
University,\\
Canberra ACT 0200, Australia}
\date{}
\maketitle



\begin{abstract}
It is shown that if the decoherence matrix corresponding to a qubit master equation has a block-diagonal real part, then the evolution is determined by a one-dimensional oscillator equation.  Further, when the full decoherence matrix is block-diagonal, then the necessary and sufficient conditions for completely positive evolution may be formulated in terms of the oscillator Hamiltonian or Lagrangian. When the solution of the oscillator equation is not known, an explicit sufficient condition for complete positivity can still be obtained, based on a Hamiltonian/Lagrangian inequality.  A rotational form-invariance property is used to characterise the evolution via a single first-order nonlinear differential equation, enabling some further exact results to be obtained.  A class of master equations is identified for which complete positivity reduces to the simpler condition of positivity.
\end{abstract}


\section{Introduction}

Master equations are useful for representing the evolution of non-isolated quantum systems, where the details of the interaction with the environment are encoded in a decoherence matrix or memory kernel \cite{breuerbook}.  In practice, given the complicated nature of typical environments, one must often work with master equations that have been derived via approximations and/or phenomenological considerations.  Unfortunately, in such cases the resulting master equations may not correspond to  evolutions that are physically possible - for example, the density operator may evolve to have negative eigenvalues.  

The distinction between physical and nonphysical master equations is not obvious in general.  For example, D\"{u}mcke and Spohn have pointed out that subtly different ways of approximating the physical principle of weak coupling, for a qubit interacting with a thermal reservoir, can variously lead to either physical or nonphysical evolution \cite{spohn}.  Similarly, Barnett and Stenholm have shown that the assumption of an apparently innocuous exponential memory kernel, describing a harmonic oscillator coupled to a reservoir, leads to negative probabilities \cite{barnett}.  

There is, therefore, interest in finding conditions on master equations which ensure that the corresponding evolution of the density operator is physical \cite{breuerbook,gorini,lindblad, lendi,ach,maniqubit,whitney,koss}. In this regard, it is not sufficient to merely ensure that the density operator remains positive under evolution.  In particular, if some auxiliary system is correlated with the system of interest, but does not interact with it, then the corresponding {\it joint} density operator must also remain positive under the evolution.  This requirement is stronger than positivity, due to the remarkable nature of quantum correlations, and is called {\it complete} positivity \cite{breuerbook,kraus,alicki}.  Determining conditions for qubit master equations to generate completely positive evolution is the focus of this paper.

The general form of a memoryless master equation, for a qubit system described by density operator $\rho$, is (eg, equation (2.7) of \cite{gorini})
\begin{equation} \label{master} 
\dot{\rho} = \Lambda_t(\rho) := - i[H(t),\rho] +(1/4)\sum_{j,k} \gamma_{jk}(t)\left( 2\,\sigma_j\rho \sigma_k - \sigma_k\sigma_j\rho - \rho\sigma_k\sigma_j \right) .
\end{equation}
Here $H(t)$ is a Hamiltonian operator, the $\sigma_j$ are the Pauli spin matrices, and $\gamma(t)$ is a $3\times 3$ Hermitian matrix which will be referred to as the decoherence matrix.  Master equations written in memory-kernel form can also be reduced to the above time-local form, provided that a particular inverse exists \cite{ach,breuer15}.  Note that the (minimal) Lindblad form of the master equation corresponds to the eigenvalue decomposition $\gamma =\sum_{l=1}^3 \lambda_l\, {\bf e}^{(l)}{\bf e}^{(l)\dagger}$ of $\gamma$, i.e., one has \cite{breuerbook,gorini,lindblad}
\[ \dot{\rho} = - i[H,\rho] +(1/4) \sum_{l=1}^3 \lambda_l \left( 2\,L_l\rho L_l^\dagger - L_l^\dagger L_l \rho - \rho  L_l^\dagger L_l \right), \]
with $L_l:={\bf e}^{(l)}\cdot\sigma $ (thus, ${\rm tr}[L_j^\dagger L_k]=2\,\delta_{jk})$.

In the case of {\it no} explicit time-dependence, the necessary and sufficient condition for completely positive evolution is simply that the decoherence matrix is nonnegative \cite{gorini, lindblad}, i.e., $\gamma \geq 0 $.  While this remains a sufficient condition when $H$ and/or $\gamma$ depend on time (since the evolution is then a composition of a sequence of infinitesimal completely positive evolutions), finding the necessary and sufficient conditions for complete positivity in the time-dependent case is a very difficult problem.  

This problem has been solved by Wonderen and Lendi \cite{lendi}, and independently by Maniscalco \cite{maniqubit}, for the case
\[ H(t) = (1/2)h\,\sigma_3,~~~~~\gamma(t)= \left(  \begin{array}{ccc} 
\gamma & ig & 0\\
-ig & \gamma & 0\\
0 & 0 & \gamma_3  
\end{array} \right) , \]
which is applicable to several systems of physical interest (including an example where the master equation is obtained from a memory-kernel form \cite{maniqubit}).  The more trivial case of characterising complete positivity when the  decoherence matrix is diagonal and $H=0$ is also solvable, and is reviewed in \cite{ach}.

In this paper the more general form
\begin{equation} \label{general}
H(t) = (1/2)h\,\sigma_3,~~~~~\gamma(t)= \left(  \begin{array}{ccc} 
\gamma_1 & f+ig & ir\\
f-ig & \gamma_2 & is\\
-ir & -is & \gamma_3  
\end{array} \right)
\end{equation}
is considered, where $f$, $g$, $h$, $r$, $s$ and the $\gamma_j$ are all real functions of time.  Thus $\gamma_{11}\neq \gamma_{22}$ in general; $\gamma_{12}$ may have a real component; and (more trivially) there is no restriction on the imaginary part of the decoherence matrix.  Note that this form is equivalent to the condition that the real part of the decoherence matrix is block-diagonal.  As will be seen in section 2, it is also equivalent to the condition that {\it the damping matrix in the Bloch representation is block-diagonal}, i.e., equation (\ref{general}) corresponds to the case that damping in one  direction is decoupled from damping in the remaining two orthogonal directions.  

It is shown in section 3 that the evolution corresponding to equation (\ref{general}) is in general determined by the solutions of a one-dimensional time-dependent oscillator equation 
\begin{equation} \label{osc}
d^2 q/d\tau^2 + k(\tau) q ,
\end{equation}
where the reparameterised time $\tau$ and the `spring constant' $k(\tau)$ are determined by $H$ and $\gamma$. Explicit solutions for the oscillator motion yield explicit solutions for the corresponding master equation.  

The general form in equation (\ref{general}) satisfies the property that the master equation is form-invariant under arbitary (time-dependent) rotations of the system about the $z$-axis.  This provides a useful gauge-like degree of freedom for simplifying the master equation, which is exploited in section 4 to characterise the qubit evolution  via a single nonlinear first-order differential equation.  This equation may be solved, for example, when 
\[  h=0,~~~~~~ f = K(\gamma_1-\gamma_2) \]
in equation (\ref{general}), for some constant $K$, which generalises the abovementioned previously considered cases \cite{lendi,ach,maniqubit}, and includes the case of a two-level atom in a time-dependent squeezed vacuum. 

In section 5 it is shown that when the {\it full} decoherence matrix is block-diagonal, i.e., when
\begin{equation}\label{block} r=s=0 \end{equation}
in equation (\ref{general}) (corresponding to any `drift' in the Bloch representation being confined to the $z$-direction), then the necessary and sufficient conditions for completely positive evolution may be formulated in terms of either the oscillator Hamiltonian or Lagrangian, depending on whether $k$ is positive or negative.  
This leads to an explicit characterisation of complete postivity, whenever the oscillator system is solvable. 

Further,  Lagrangian and Hamiltonian oscillator inequalities are derived which lead to a useful sufficient condition for complete positivity.  In particular, this condition is formulated solely in terms of the master equation parameters, and hence can be applied whether or not the corresponding solution is known.  It is quite strong, being both necessary and sufficient in a number of cases, and is invariant under rotations of the system about the $z$-axis.  

Finally, a class of master equations is identified for which complete positivity is equivalent to positivity, corresponding to the case of unital evolution with no damping in the $z$-direction.  This class includes a group of Bloch-Redfield master equations in a white noise limit, recently considered by Whitney, which generate positive evolution \cite{whitney}.  It immediately follows that the evolution is in fact completely positive for this group.  This case is of further interest in that the question of complete positivity can be settled despite being unable to solve for the evolution explicitly.

\section{Evolution in the Bloch representation}

It is convenient to rewrite the master equation (\ref{master}) in terms of the Bloch vector ${\bf v}$, where
\begin{equation} \label{bloch} 
\rho = (1/2)\,[1 + {\bf v}\cdot\sigma].  \end{equation}
Substituting into the first equality in (\ref{master}) and taking the trace with $\sigma_j$ then leads to the equivalent Bloch equation 
\[  \dot{\bf v} = {\bf u} + D{\bf v} , \]
where 
\[ u_j := (1/2)\,{\rm tr}[\sigma_j\Lambda(1)], ~~~~~~D_{jk}:= (1/2)\,{\rm tr}[\sigma_j\Lambda(\sigma_k)]  \]
are termed the drift vector and damping matrix respectively \cite{maniqubit}. This is a first-order inhomogenous differential equation, and hence the general solution is of the form
\begin{equation} \label{blochsol} 
{\bf v}(t) =  M(t)\,{\bf v}(0) + {\bf w}(t) , 
\end{equation}
for some matrix $M$ and vector ${\bf w}$.  
Substitution of (\ref{blochsol}) into the Bloch equation yields the equivalent evolution equations
\[ \dot{M} = DM,~~~~~\dot{\bf w} = {\bf u} + D{\bf w}, \]
for $M$ and $\bf{w}$, subject to the inital conditions 
\[  M(0)=I,~~~~~ {\bf w}(0) = {\bf 0} .  \]

Now, to first-order in $\epsilon$, 
\[ \det M(t+\epsilon) = \det M \det(I+\epsilon D) = \det M \prod_j (1+\epsilon D_{jj}) = \det M (1+\epsilon{\rm tr}[D]), \]
and hence it follows that
\[ \det M = \exp \left( \int_0^t ds \, {\rm tr}[D(s)] \right) >0 .  \]
Thus the inverse matrix $M^{-1}$ always exists, and it is straightforward to check that the solution for ${\bf w}$ is  given by
\begin{equation} \label{wsol}
{\bf w}(t) = M(t) \int_0^t ds\, M^{-1}(s)\,{\bf u}(s) .
\end{equation} 
Hence, solving the general master equation (\ref{master}) is equivalent to solving the matrix equation 
\begin{equation} \label{mdot}
\dot{M}=DM,~~~~~ M(0)=I 
\end{equation}
for the evolution matrix $M$.  Similar considerations apply to higher-dimensional quantum systems.

For the general qubit master equation  in (\ref{master}), the above definitions of ${\bf u}$ and $D$ and the relations
\begin{equation} \label{sigma}
{\rm tr} [\sigma_a\sigma_b\sigma_c] = 2i\epsilon_{abc},~~~~~
{\rm tr} [\sigma_a\sigma_b\sigma_c\sigma_d] = 2\left(\delta_{ab}\delta_{cd} + \delta_{ad} \delta_{bc}  - \delta_{ac} \delta_{bd}     \right) 
\end{equation}
may be used to calculate
\[ u_1 = -{\rm Im}\{\gamma_{23}\}~~{\rm et~cyclic},~~~~~D_{jk} = {\rm Re}\{\gamma_{jk}\} -\delta_{jk}\, {\rm tr}[\gamma] -\sum_l \epsilon_{jkl}\,{\rm tr}[H\sigma_l]  . \]
It follows that, for the particular forms of $H$ and $\gamma$ defined in (\ref{general}), the drift vector and damping matrix are given by
\begin{equation} \label{drift}
{\bf u} = \left( \begin{array}{c} -s\\r\\-g \end{array} \right) ,~~~~~
D = \left( \begin{array}{ccc} 
-\gamma_2-\gamma_3 & f-h & 0\\
f+h & -\gamma_1-\gamma_3 & 0\\
0 & 0 & -\gamma_1 - \gamma_2 \end{array} \right) .
\end{equation}
It is the block-diagonal form of the damping matrix $D$, corresponding to damping in the $z$-direction being decoupled from damping in the $x$ and $y$ directions, that provides the basis for the main results of this paper.
  
\section{Reduction to an oscillator system}

\subsection{Oscillator form}

From equations (\ref{mdot}) and (\ref{drift}), the evolution matrix $M$ is itself block-diagonal, i.e., one has
\begin{equation} \label{mform}
M = \left( \begin{array}{ccc}
x_1 & x_2 & 0\\
y_1 & y_2 & 0\\
0 & 0 & A \end{array} \right). 
\end{equation}
Substitution into the evolution equation (\ref{mdot}) yields in particular that
\[ dA/dt = -(\gamma_1+\gamma_2)A,~~~~~d \Delta/dt= -(\gamma_1 + \gamma_2 + 2\gamma_3)\, \Delta, \]
where $\Delta$ denotes the subdeterminant $x_1y_2-x_2y_1$, and thus
\begin{equation} \label{a}
A = \exp \left[-\int_0^t ds\,(\gamma_1+\gamma_2)\right], ~~~~~\Delta = A \exp \left[-2 \int_0^t ds\, \gamma_3\right] . 
\end{equation}
Hence only three independent parameters of $M$ remain to be determined.

Consider now the 2-vector equation
\[ \left( \begin{array}{c} \dot{x}\\ \dot{y} \end{array} \right) = 
\left( \begin{array}{cc} -\gamma_2-\gamma_3 & f-h\\
f+h & -\gamma_1-\gamma_3 \end{array} \right)\,
\left( \begin{array}{c} x\\y \end{array} \right) , \]
where $(x,y)=(x_j,y_j)$ ($j=1$, $2$), corresponding to the evolution of the upper block of $M$.  Defining the quantities
\begin{equation} \label{qpg} 
q:=x e^{ \int_0^t ds\,(\gamma_2+\gamma_3) },~~~~~p:=  e^{ \int_0^t ds\,(\gamma_1+\gamma_3)}y ,~~~~~ G:= e^{\int_0^t ds \,(\gamma_1-\gamma_2)}, 
\end{equation}
one immediately finds that 
\[ \dot{q} = (f-h)G^{-1}p,~~~~~ \dot{p} = (f+h) Gq ,\]
and hence the evolution of $q$ and $p$ is described by the quadratic Hamiltonian
\[ H(q,p,t) := \frac{1}{2} (f-h)G^{-1}p^2 - \frac{1}{2}(f+h)G q^2  . \]

To obtain the canonical oscillator form, note that the corresponding action, $\int dt\,(p\dot{q}-H)$, may be rewritten as
\[ \frac{1}{2} \int dt\,\left[ (f-h)^{-1}G\dot{q}^2 + (f+h)Gq^2\right] = \frac{1}{2}\int d\tau\,\left[ \left(\frac{dq}{d\tau}\right)^2 - kq^2 \right] , \]
providing that one defines the functions $\tau(t)$ and $k(\tau)$ via 
\begin{equation} \label{tau} 
\dot{\tau}:= (f-h)G^{-1},~~~~~\tau(0)=0,~~~~~k:= G^2(h+f)/(h-f)  .
\end{equation}
Hence, with respect to the reparameterised time $\tau$, one has the oscillator equation 
\begin{equation} \label{osc2}
d^2 q/d\tau^2 + k q =0 ,  
\end{equation}
as previewed in the introduction. 

It follows that when the oscillator equation (\ref{osc2}) can be solved, then the evolution matrix $M$ can be determined.  In particular, such a solution must link the oscillator state to its inital state via a linear relation of the form
\begin{equation} \label{link}
\left( \begin{array}{c} q\\ dq/d\tau \end{array} \right) = 
\left( \begin{array}{cc} a & b\\
c & d \end{array} \right)\,
\left( \begin{array}{c} q_0\\(dq/d\tau)_0 \end{array} \right) . 
\end{equation}
Since $dq/d\tau=\dot{q}/\dot{\tau}=p$ from the defining equations (\ref{qpg}) and (\ref{tau}), it follows further via (\ref{qpg}) and (\ref{link}) that 
\[ \left( \begin{array}{c} x\\ y \end{array} \right) = 
\left( \begin{array}{cc} e^{-\int ds\,(\gamma_2+\gamma_3)} & 0\\
0 & e^{-\int ds\,(\gamma_1+\gamma_3)} \end{array} \right)\,
\left( \begin{array}{cc} a & b\\
c & d \end{array} \right)\,
\left( \begin{array}{c} x_0\\y_0 \end{array} \right) . \]
Recalling the initial condition $M(0)=I$, the corresponding explicit form
\begin{equation} \label{mfinal}
M = \left( \begin{array}{ccc} 
a\,e^{-\int ds\,(\gamma_2+\gamma_3)} & b\,e^{-\int ds\,(\gamma_2+\gamma_3)} & 0\\
c\,e^{-\int ds\,(\gamma_1+\gamma_3)} & d\,e^{-\int ds\,(\gamma_1+\gamma_3)} & 0 \\
0& 0 & e^{-\int ds\,(\gamma_1+\gamma_2)}
\end{array} \right)
\end{equation}
is obtained for the evolution matrix.  Thus, the master equation is solvable whenever the oscillator matrix in (\ref{link}) is known.

\subsection{Examples}

As a simple example, consider the case $f=0=\gamma_1-\gamma_2$.  Then, $k=G= 1$ and $\tau = -\int_0^t ds\,h(s)$ from equations (\ref{qpg}) and (\ref{tau}), and hence from (\ref{osc2}) the oscillator matrix is
\[ \left( \begin{array}{cc} a & b\\
c & d \end{array} \right) = \left( \begin{array}{cc} \cos \tau & \sin \tau \\
\ -\sin \tau & \cos \tau \end{array} \right) = \left( \begin{array}{cc} \cos \int_0^t ds\,h(s) & -\sin \int_0^t ds\,h(s)\\
\sin \int_0^t ds\,h(s) & \cos \int_0^t ds\,h(s) \end{array} \right) .  \]
The solution of the corresponding master equation in the Bloch representation then follows via equations (\ref{bloch})-(\ref{wsol}) and (\ref{mfinal}).  Note this case corresponds to that considered previously by Wonderen and Lendi \cite{lendi} and Maniscalco \cite{maniqubit}.  It is significantly generalised in the following section.

As a second example, consider the case $h=0=\gamma_1-\gamma_2$, corresponding to a symmetric damping matrix in the interaction picture.  One finds $k= -1$, $G=1$ and $\tau=\int_0^t ds\,f(s)$ and hence that
\[ \left( \begin{array}{cc} a & b\\
c & d \end{array} \right) = \left( \begin{array}{cc} \cosh \int_0^t ds\,f(s) & \sinh \int_0^t ds\,f(s)\\
\sinh \int_0^t ds\,f(s) & \cosh \int_0^t ds\,f(s) \end{array} \right) .  \]

Finally, it proves useful to consider the degenerate case $f=-h$, for which one has $k\equiv 0$ and $\tau\equiv 2\int_0^t ds\,fG^{-1}$.  Thus the corresponding oscillator system degenerates to free particle motion, with zero frequency, and
\[  \left( \begin{array}{cc} a & b\\
c & d \end{array} \right)  =  \left( \begin{array}{cc} 1 & \tau\\
0 & 1 \end{array} \right)=  \left( \begin{array}{cc} 1 & 2\int_0^t ds\, f\,e^{\int_0^s du \,(\gamma_2-\gamma_1)}\\
0 & 1 \end{array} \right) .  \]
Note that this reduces to the identity matrix when one further has $f=0$, corresponding to the trivial case of a diagonal damping matrix in equation (\ref{drift}) (previously reviewed in \cite{ach}). More significantly, this solution is used in the following section, together with rotational form-invariance, to further reduce the evolution of the master equation to a single nonlinear {\it first-order} equation. 

Note for all the above examples that 
\begin{equation} \label{det}
ad-bc = 1  .
\end{equation}
This holds more generally, as may be derived directly from the oscillator equation (\ref{osc2}), or via equations (\ref{a}) and (\ref{mfinal}).  Note also that the oscillator equation breaks down for the singular case $f=h$, since $k$ becomes undefined in equation (\ref{tau}).  However, one can solve equation (\ref{mdot}) for $M$ directly in this case, to find it corresponds to an oscillator matrix equal to the transpose of the one given above for $f=-h$.

\section{Exploiting rotational form-invariance}

\subsection{Rotational form-invariance}

Consider now a rotation of the system about the $z$-axis, by some time-dependent angle $\alpha(t)$.  Such a rotation corresponds to the unitary transformation $\tilde{\rho}:= U\rho U^\dagger $ of the density operator, with $U := \exp[-i\alpha(t)\sigma_3]$.  Denoting the corresponding rotation matrix by $R$, one has $U^\dagger\sigma_j U = \sum_k R_{jk}\sigma_k$, and substitution into (\ref{master}) yields the transformed master equation
\[ \dot{\tilde{\rho}} = \tilde{\Lambda}_t(\tilde{\rho}) := - i[\tilde{H}(t),\tilde{\rho}] +(1/4)\sum_{j,k} \tilde{\gamma}_{jk}(t)\left( 2\,\sigma_j\tilde{\rho} \sigma_k - \sigma_k\sigma_j\tilde{\rho} - \tilde{\rho} \sigma_k\sigma_j \right) , \]
with
\[ \tilde{H} = H + \dot{\alpha}\,\sigma_3,~~~~~\tilde{\gamma} = R\gamma R^T . \]
A tilde will be used in general to denote quantities in the rotated frame.

For the particular class of `Bloch-diagonal' master equations described by equation (\ref{general}), one finds in particular that
\begin{equation} \label{rotate}
\tilde{H} = (1/2)\tilde{h}\, \sigma_3,~~~~~~\tilde{\gamma} = \left(  \begin{array}{ccc} 
\tilde{\gamma}_1 & \tilde{f}+ig & i\tilde{r}\\
\tilde{f}-ig & \tilde{\gamma}_2 & i\tilde{s}\\
-i\tilde{r} & -i\tilde{s} & \gamma_3  
\end{array} \right) ,
\end{equation}
where
\[ \tilde{\gamma}_1 := \gamma_1\cos^2 \alpha+\gamma_2\sin^2\alpha-f\sin 2\alpha,~~~~~\tilde{\gamma}_2 :=\gamma_1\sin^2\alpha+\gamma_2\cos^2\alpha+f\sin 2\alpha, \]
\[ \tilde{h} = h+2\dot{\alpha},~~~~~\tilde{f} = f\cos 2\alpha +(1/2)(\gamma_1-\gamma_2)\sin 2\alpha  , \] 
\[ \tilde{r}= r\cos\alpha -s\sin\alpha,~~~~~\tilde{s} = r\sin\alpha + s\cos\alpha .  \]
Note that $g$, $\gamma_1+\gamma_2$ and $\gamma_3$ do not change under the rotation.

Comparing equations (\ref{general}) and (\ref{rotate}) it is seen that the form of equation (\ref{general}) is preserved by such rotations.    Moreover, from equation (\ref{bloch}) and the property $U^\dagger\sigma U = R\sigma$, the Bloch vector transforms under such rotations as
\[ \tilde{{\bf v}} = R{\bf v} .\]
Substitution into equation (\ref{blochsol}) then yields the relations
\begin{equation} \label{mtilde} 
M = R^T\tilde{M}R(0),~~~~~{\bf w} = R^T\tilde{\bf w}  .  
\end{equation}
Thus, if the master equation can be solved in the rotated system, by a judicious simplifying choice of the function $\alpha(t)$, then the solution with respect to the original system can also be determined.  This rotational degree of freedom allows non-trival new exact solutions to be obtained for qubit master equations, via reduction of the evolution to a single first-order differential equation for $\alpha$, as is shown below.  It is also relevant to the discussion of complete positivity in section 5, where it is used to obtain a rotationally-invariant sufficient condition.

\subsection{Reduction to a first-order nonlinear equation}

For any master equation of the form (\ref{general}), define the `zero frequency' gauge or picture via the condition
\[ \tilde{f} = - \tilde{h}.\]
Note that this is rather different from the standard `interaction' picture, which corresponds to $\tilde{h}=0$.  From the above expressions for  $\tilde{f}$ and $\tilde{h}$, this condition may be rewritten as the first order differential equation
\begin{equation} \label{gauge}
\dot{\alpha} + \frac{1}{4} \left[ 2f\cos 2\alpha + (\gamma_1-\gamma_2)\sin 2\alpha + 2h \right] = 0 
\end{equation}
for $\alpha$.  

Now, the choice $\tilde{f} = - \tilde{h}$ corresponds to the degenerate case of a zero-frequency oscillator considered in section 3.2, and therefore the corresponding oscillator matrix can be immediately written down as
\begin{equation} \label{oscmat}  
\left( \begin{array}{cc} \tilde{a} & \tilde{b}\\
\tilde{c} & \tilde{d} \end{array} \right)  =  \left( \begin{array}{cc} 1 & 2\int_0^t ds\, \tilde{f}\,e^{\int_0^s du \,(\tilde{\gamma_2}-\tilde{\gamma_1})}\\
0 & 1 \end{array} \right) .  
\end{equation}
Hence, if $\alpha$ can be determined from equation (\ref{gauge}), then $\tilde{M}$ can be determined via (\ref{mfinal}), and the evolution matrix $M$ for the original master equation follows via (\ref{mtilde}).

Thus, remarkably, solving the master equation is equivalent to solving the zero-frequency gauge equation (\ref{gauge}) for $\alpha$. 
Naturally enough, solving this equation explicitly cannot be done in general, as it would amount to solving a general time-dependent oscillator problem. 

\subsection{Example: a new solution}

It is possible to solve equation (\ref{gauge}) in some cases of interest other than the examples of the previous section.  For example, consider the case 
\begin{equation} \label{special}
h=0,~~~~~~ f = K(\gamma_1-\gamma_2)
\end{equation}
for some constant $K$.  This case corresponds to $k(\tau)$ being explicitly time-dependent in the oscillator equation (\ref{osc2}), and includes the known cases $\gamma_1-\gamma_2=f=h=0$ \cite{lendi,maniqubit} and $f=h=0$ \cite{ach} in particular. However, it is rather more general, including, for example, a two-level atom with natural linewidth $\gamma$ coupled to a squeezed vacuum described by squeezing parameter $\xi(t)=r(t)e^{i\theta_0t}$, with (slowly varying) time-dependent squeezing parameter $r(t)$ and fixed squeezing angle $\theta_0$. This corresponds to the particular choice \cite{breuerbook}
\[ \gamma_1+\gamma_2=\gamma\cosh 2r,~~~~\gamma_1-\gamma_2=-\gamma\cos\theta_0 \sinh 2r,~~~~f=-(1/2) \gamma\sin\theta_0 \sinh 2r  \]
in equation (\ref{general}), with all other coefficients vanishing, and hence to $K=-\frac{1}{2}{\rm cot}~\theta_0$ in equation (\ref{special}) above.

To solve the master equation in this case, define the (constant) angle $\phi$ by
\[\cos 2\phi := 2K/\sqrt{1+4K^2},~~~~~\sin 2\phi:= 1/\sqrt{1+4K^2}. \]
The `zero frequency' equation (\ref{gauge}) can then be rewritten in the separable form
\[ (\sin 2\phi)\, \frac{d}{dt}\, 2(\alpha -\phi) = -(1/2) (\gamma_1-\gamma_2) \cos 2(\alpha-\phi) , \]
which may be immediately be integrated to give, assuming that $\alpha(0)=\phi$ for convenience,
\[ (\sin 2\phi) \log \tan (\alpha-\phi+\pi/4) = -(1/2) \int_0^t ds\,(\gamma_1-\gamma_2) .  \]
Inverting gives the explicit expression
\begin{equation} \label{alpha}
\alpha = \phi - \pi/4 + \tan^{-1}\left[ e^{-(1/4K)\sqrt{1+4K^2}\int_0^t ds\, (\gamma_1-\gamma_2) }
  \right]  
\end{equation}
for the gauge function $\alpha$, as desired.  Thus, for all master equations satisfying condition (\ref{special}), the explicit evolution in the Bloch representation can be obtained by (i) applying equation (\ref{mfinal}) to equation (\ref{oscmat}) for the above choice of $\alpha$, to obtain $\tilde{M}$ , and (ii) finding the evolution matrix $M$ via relation (\ref{mtilde}).

\section{Complete positivity for block-diagonal $\gamma$}

\subsection{Necessary and sufficient conditions}

The evolution of a quantum system, described by some linear map $\rho_t=\phi_t(\rho_0)$, is completely positive if and only if the corresponding Choi matrix $C$ has no negative eigenvalues, i.e., if and only if $C\geq 0$ \cite{choi}.  For the case of qubits, it is convenient to calculate this matrix $C$ with respect to the basis set used in section 4 of reference \cite{ach} (corresponding to the matrix $S^{(W)}$ therein), and multiply by a factor of 2, so that the Choi matrix is the $4\times 4$ matrix defined by
\[ C_{jk} := (1/2) \sum_{m,n} F_{mn}\,{\rm tr} [\sigma_n \sigma_j \sigma_m \sigma_k] , \]
where $F_{mn}:=(1/2){\rm tr}[\sigma_m\phi(\sigma_n)]$, and the indices run over $0$, $1$, $2$, $3$ with $\sigma_0:=1$.

From the Bloch representation in equations (\ref{bloch}) and (\ref{blochsol}), the evolution map $\phi$ is given by
\[ \phi(X) = (1/2){\rm tr}[X]\, \left(1+{\bf w}\cdot\sigma \right) +(1/2) \left(M{\rm tr}[X\sigma]\right)\cdot\sigma \]
where $\sigma$ denotes the $3$-vector $(\sigma_1,\sigma_2,\sigma_3)$, and hence $F_{00}=1$, $F_{0j}=\delta_{j0}$, $F_{j0}=w_j$, and $F_{jk}=M_{jk}$ for $j,k=1,2,3$.  It follows, using properties (\ref{sigma}) of the Pauli matrices, that the coefficients of the Choi matrix are given by
\[ C_{00}=1+{\rm tr}[M],~~~~C_{01}= w_1 +i(M_{23}-M_{32}),\]
\[ C_{11}=1+M_{11}-M_{22}-M_{33},~~~~C_{12} = M_{12}+M_{21}+iw_3 , \] 
with the remaining coefficients determined via cylic permutations of $1,2,3$ and $C=C^\dagger$.
Checking positivity of the Choi matrix for a general qubit evolution essentially requires finding the singular values of $M$ \cite{ruskai}, and hence there is no general explicit condition in terms of the coefficients of $M$ and $\bf{w}$.  

However, for master equations with a block-diagonal decoherence matrix, the Choi matrix has a relatively simple form.  This case is equivalent to $r=s=0$ in equation (\ref{general}) and implies that, in addition to the damping matrix $D$ being diagonal, the `drift' vector ${\bf u}$ in (\ref{drift}) is confined to the $z$-direction.  Equations (\ref{wsol}), (\ref{drift}) and (\ref{mform}) then lead to  
\begin{equation} \label{choi}
C = \left( \begin{array}{cccc}
1+x_1+y_2+A & 0 & 0 & w_3 + i (x_2-y_1)\\
0 & 1+x_1 - y_2 - A & x_2+y_1 + iw_3 & 0\\
0  & x_2+y_1-iw_3 & 1 -x_1+y_2 - A & 0\\
w_3 - i(x_2-y_1)& 0 & 0 & 1-x_1-y_2 +A
\end{array} \right)  .
\end{equation}
for the corresponding Choi matrix, with 
\begin{equation} \label{w3} 
w_3 = -A\int_0^t ds\,gA^{-1} . 
\end{equation}

The condition $C\geq 0$ thus reduces to the positivity of the two $2\times 2$ submatrices composing $C$, i.e., to the positivity of the traces and determinants of these submatrices.  This yields, after some rearrangement, the necessary and sufficient conditions 
\begin{equation} \label{cp} 
A\leq 1,~~~~~ S:=x_1^2+x_2^2+y_1^2+y_2^2\leq 1+A^2 -w_3^2 -2|A-\Delta| 
\end{equation}
for complete positivity.  Note from (\ref{a}) that $A$ and $\Delta$ are explicitly defined in terms of the master equation parameters, as is $w_3$ (given above).

The question of complete positivity therefore reduces to knowledge about the quantity $S$ on the left hand side of second inequality in (\ref{cp}).  This quantity may of course be calculated when the solution of the master equation is known, such as for the examples in sections 3 and 4, thus completely determining the conditions for complete positivity in these cases.  More generally, however, only partial conditions can be explicitly determined in terms of the master equation parameters, as discussed below.

\subsection{Necessary conditions}

Here two {\it necessary} conditions for complete positivity are noted, for master equations having a block-diagonal decoherence matrix, which do not require the solution of the master equation.  Both conditions are formulated in terms of quantities that are invariant under rotations about the $z$-axis.

First, the condition $A\leq 1$ in (\ref{cp}) reduces via equation (\ref{a}) to
\[ \int_0^t ds\, (\gamma_1 + \gamma_2) \geq 0 . \]

Second, noting that the quantity $S$ in (\ref{cp}) is the sum of the squares of the singular values of the upper block of $M$ in (\ref{mform}), and that the positive quantity $\Delta$ is their product, it follows via $s_1^2+s^2\geq 2s_1s_2$ and (\ref{a}) that complete positivity requires
\[ A^2 - 2A\left[ | 1-e^{-2\int_0^t ds\,\gamma_3} | + e^{-2\int_0^t ds\,\gamma_3}\right]+w_3^2+1\geq 0 . \]
Note that this quadratic condition is certainly satisfied when the  corresponding discriminant is negative, i.e., when
\[ | 1-e^{-2\int_0^t ds\,\gamma_3} | + e^{-2\int_0^t ds\,\gamma_3} \leq \sqrt{1+w_3^2} , \]
which in turn is guaranteed when
\[ \int_0^t ds\,\gamma_3 \geq 0  . \]

\subsection{Sufficient condition from a Lagrangian inequality}

To obtain a nontrivial sufficient condition for complete positivity, it is convenient to begin by working in the interaction picture, so that $h=0$.  In this case one has $k=-G^2$ for  the corresponding (inverted) oscillator system in section 3, and hence the oscillator Lagrangian is given by
\[ L(q,dq/d\tau,\tau) = (1/2)\left[(dq/d\tau)^2 + G^2 q^2 \right] .\]
Noting that the conjugate momentum is $p=dq/d\tau$, the value of the Lagrangian at any given time follows from equation (\ref{link}) as
\[  L = (1/2)G^2 (aq_0 + bp_0)^2 + (1/2) (cq_0+dp_0)^2 . \]
Hence, if $L_1$ and $L_2$ refer to the values of $L$ at time $t$ for the canonical initial states $(q_0,p_0)=(1,0)$ and  $(q_0,p_0)=(0,1)$ respectively (actually, any two orthogonal initial states of equal norm will do), then their {\it average} value evolves as
\[ \overline{L} := (1/2)(L_1 + L_2) = (1/2)\left[G^2 (a^2 + b^2) +  (c^2+d^2) \right] .  \]
Note that $\overline{L}$ is invariant under phase space rotations.

Comparing this expression with equations (\ref{mform}) and (\ref{mfinal}), and noting the definition of $G$ in equation (\ref{qpg}), it follows that
\begin{equation} \label{j}
S = x_1^2+x_2^2+y_1^2+y_2^2 = 2 \overline{L} e^{-2\int_0^t ds\,(\gamma_1+\gamma_3)} .
\end{equation}
Hence {\it the complete positivity condition (\ref{cp}) may be interpreted as an upper bound on the average Lagrangian value of the corresponding oscillator system}.

In particular, any upper bound for $\overline{L}$ immediately generates a {\it sufficient} condition for complete positivity.  One such bound is obtained here, using a generalisation of the method given by Boonserm and Visser for obtaining  bounds for $a^2+b^2+c^2+d^2$ (rather than for $\overline{L}$), relevant to one-dimensional scattering coefficients \cite{visser}.  In section 5.3 this bound is shown to in fact be applicable to the case of arbitrary $h$, as a consequence of rotational invariance.  This further allows a Hamiltonian upper bound to be obtained
for the time-dependent harmonic oscillator.

First, define the quantities $X_{\pm}$, $Z$ by
\[ X_{\pm} := G (a^2+b^2) \pm G^{-1}(c^2+d^2),~~~~~Z:=ac+bd  .\]
Note from the determinant property (\ref{det}) that
\[ X_+^2 - X_-^2 = 4(a^2+b^2)(c^2+d^2) = 4(ac+bd)^2+4(ad-bc)^2=4Z^2+4 .\]
Now, the oscillator equations (\ref{osc2}) and (\ref{link}) imply that
\[ \frac{d}{d\tau} \left( \begin{array}{cc}
a & b\\
c & d \end{array} \right) = 
\left( \begin{array}{cc}
0 & 1\\
G^2 & 0 \end{array} \right)
\left( \begin{array}{cc}
a & b\\
c & d \end{array} \right) , \]
from which it follows, writing $G'=dG/d\tau$, that
\[ dX_+/d\tau = (G'/G)X_- +4GZ = (G'/G,2G)\cdot (X_-,2Z) . \]
But $d/d\tau\equiv fG^{-1}(d/dt)$ from equation (\ref{tau}) and hence, noting the definition of $G$ in (\ref{qpg}) and making use of the Schwarz inequality, one finds
\[ \dot{X}_+ = (\gamma_1-\gamma_2,2f)\cdot (X_-,2Z) \leq [(\gamma_1-\gamma_2)^2+4f^2]^{1/2}[X_-^2+4Z^2]^{1/2} .\]
Combining this result with the above expression for $X_+^2-X_-^2$ then yields 
\[ [X_+^2-4]^{-1/2}\,\dot{X}_+ \leq [(\gamma_1-\gamma_2)^2+4f^2]^{1/2} ,\]
which may be integrated to give
\[ \cosh^{-1} X_+/2 \leq \int_0^t ds\, [(\gamma_1-\gamma_2)^2+4f^2]^{1/2} . \]
Noting $\overline{L} = GX_+/2$, one finally obtains the Lagrangian inequality
\begin{equation} \label{ineq}
\overline{L} \leq  G \cosh \left[ \int_0^t ds\, [(\gamma_1-\gamma_2)^2+4f^2]^{1/2}  \right] .
\end{equation}

A sufficient condition for complete positivity in the interaction picture follows immediately from (\ref{a}), (\ref{cp}), (\ref{j}) and (\ref{ineq}) as 
\begin{equation} \label{bound}
A\leq 1,~~~~ \Delta \cosh \left[ \int_0^t ds\, [(\gamma_1-\gamma_2)^2+4f^2]^{1/2}  \right] \leq 1+A^2 -w_3^2 -2|A-\Delta| .
\end{equation}
It is important to emphasise that this condition can be checked {\it whether or not} the master equation can be explicitly solved, as it depends only on the decoherence matrix $\gamma$ in the interaction picture.  

Note that the above condition is `tight' in the sense that it is in fact {\it necessary} and sufficient in some cases.  For example, when $\gamma_1-\gamma_2=f=0$ in the interaction picture \cite{lendi,maniqubit}, then $G\equiv 1$, $a=d=1$, and $b=c=0$, implying that equality holds in (\ref{ineq}).  It is in fact also `universal', i.e., it is valid for $h\neq 0$ as well as for $h=0$, as will be shown below.

\subsection{Generalisations via form-invariance}

While the sufficient condition (\ref{bound}) for complete positivity was only derived for the interaction picture, with $h=0$, it is in fact invariant under rotations about the $z$-axis, and hence may be applied to any master equation with a block-diagonal decoherence matrix.

In particular, for the general case where $h$ is an arbitrary function of time, consider a rotation about the $z$-axis such that $\tilde{h}=0$, corresponding to the choice $\alpha(t)=-(1/2)\int_0^t ds\,h$ (see section 4.1).  Due to the rotational form-invariance of the master equation, it follows that condition (\ref{bound}) must hold with respect to the associated decoherence matrix $\tilde{\gamma}$, i.e., one has the sufficient condition
\[ \tilde{A}\leq 1,~~~~ \tilde{\Delta} \cosh \left[ \int_0^t ds\, [(\tilde{\gamma_1}-\tilde{\gamma_2})^2+4\tilde{f}^2]^{1/2}  \right] \leq 1+\tilde{A}^2 -\tilde{w}_3^2 -2|\tilde{A}-\tilde{\Delta}|  \]
for complete positivity.  However, using the tranformation equations for $\gamma_1$, $\gamma_2$, etc. in section 4.1, one finds that all the relevant quantities are rotationally invariant.  In particular,one has $\tilde{A}=A$, $\tilde{\Delta}=\Delta$, $\tilde{w}_3=w_3$, and $(\tilde{\gamma_1}-\tilde{\gamma_2})^2+4\tilde{f}^2 = (\gamma_1-\gamma_2)^2+4f^2$ for {\it any} choice of $\alpha$.  Hence, condition (\ref{bound}) is in fact universal.

The universal form of (\ref{bound}) is a fortunate consequence of choosing to work in the interaction picture in section 5.3.  If, for example, $f=0$ instead of $h=0$ had been assumed, then one would have obtained the simpler evolution $dX_+/d\tau = (G'/G)X_-$, leading to the {\it Hamiltonian} inequality 
\[ \overline{H} \leq G \cosh \left[\int_0^t ds\, |\gamma_1-\gamma_2|\right]  \]
analogous to the Lagrangian inequality (\ref{ineq}), where $\overline{H}$ replaces $\overline{L}$ in (\ref{j}).  Thus the integrand in equation (\ref{bound}) would have been replaced by the quantity $|\gamma_1-\gamma_2|$, which is clearly {\it not} universal.  However, the rotationally invariant form in (\ref{bound}) can then still be obtained, by considering an arbitrary rotation about the $z$-axis.

Finally, recalling that $X_+^2 = X_-^2 +Z^2 +4\geq 4$, it is worth noting that the Lagrangian inequality (\ref{ineq}) may be extended to 
\begin{equation}  1 \leq \overline{L}/\omega \leq  \cosh \left[\log \omega_0+ \int_0^t ds\, [(\dot{\omega}/\omega)^2 +4\omega^2 ]^{1/2}\right] \end{equation}
for the average Lagrangian value of the general inverted oscillator equation $\ddot{x}-\omega^2x=0$.  Similarly, the above Hamiltonian inequality may be extended to 
\begin{equation} 1 \leq \overline{H}/\omega \leq  \cosh \left[ \log\omega_0+\int_0^t ds\, |\dot{\omega}/\omega| \right] \end{equation}
for the average Hamiltonian value of the general oscillator equation $\ddot{x}+\omega^2x=0$.   The upper bounds in these equations can presumably be generalised to include an arbitrary function $\Omega$, analogous to those for $a^2+b^2+c^2+d^2$ in \cite{visser}, leading to new sufficient conditions for complete positivity.  Minimisation with respect to the choice of $\Omega$ would then give a `best' sufficient condition.  However, this will not be further investigated here.

\subsection{Complete positivity for $xy$-damping and zero drift}

The condition that the density matrix remains a positive operator under evolution is generally weaker than the requirement of complete positivity, as remarked in the introduction.  For qubits, positivity is equivalent to the requirement that the Bloch vector remains in the unit ball, i.e., $|{\bf v}| \leq 1$.  For unital evolution, corresponding to the case of a {\it real} decoherence matrix $\gamma$, and hence ${\bf w}={\bf 0}$ in equation (\ref{blochsol}),  positivity is therefore equivalent to the condition
\[  M^TM \leq I  .\]

Consider now a master equation with a decoherence matrix that is both real {\it and} block-diagonal, i.e., with $g=r=s=0$ in equation (\ref{general}).   It will further be assumed that $\gamma_3=0$.  These conditions correspond in the Bloch representation to  zero drift and to the confinement of any damping to the  $xy$-plane.  After some minor algebra, the condition for positivity reduces in this case to
\begin{equation} A \leq 1,~~~~S=x_1^2+x_2^2+y_1^2 +y_2^2 \leq 1+A^2  .
\end{equation}
Noting from (\ref{wsol}) and (\ref{a}) that $A=\Delta$  and $w_3=0$, this condition is in fact equivalent to (\ref{cp}).  
Hence, complete positivity reduces to the generally weaker property of positivity for this particular class of master equations.  

An interesting example of this equivalence is provided by a Bloch-Redfield master equation recently studied by Whitney \cite{whitney}, corresponding to a two-level system coupled via $\sigma_1$ to a thermal environment, in the limit of a short memory time and high temperature (see equation (22) of \cite{whitney}).  Although this master equation cannot be solved explicitly, Whitney has shown via asymptotic analysis of $|{\bf v}|$ that the corresponding evolution satisfies positivity.  It follows immediately from the above that the evolution must therefore in fact be completely positive, despite being unable to construct, for example, an explicit Kraus representation. 

\section{Conclusions}

The main results of this paper are the reduction of a large class of master equations to a time-dependent oscillator system and further to a nonlinear first-order differential equation, in sections 3 and 4 respectively; the generation of new exact solutions using these reductions and rotational form-invariance; and the characterisation of complete positivity in terms of an oscillator Lagrangian/Hamiltonian, leading to a rotationally-invariant sufficient condition for complete positivity in section 5.  

Note that further examples of exact solutions, for master equations having decoherence matrices as per equation (\ref{general}), can of course be generated from any exactly solvable oscillator equation.  For example, suppose the oscillator equation (\ref{osc2}) can be solved for some invertible `spring constant' function $k(\tau)$ with inverse $k^{-1}(\tau)$, satisfying $k(0)=1$ (which can be ensured by translating and/or rescaling $\tau$).  Making the anzatz $f=0$,  it follows immediately from (\ref{tau}) that $\dot{\tau} =-h/G = (d/dt)[k^{-1}(G^2)]$.  Hence, using the procedure in section 3, one can solve the master equation corresponding to the case
\[ f=0,~~~~h=-G(d/dt)[k^{-1}(G^2)]  \]
in equation (\ref{general}), for arbitrary $\gamma_1$ and $\gamma_2$, where $G$ is defined in equation (\ref{qpg}).  Thus any single oscillator solution generates solutions for a large class of master equations.  Moreover, one can also obtain solutions for the case $f\neq 0$ by applying the above result to the rotated system defined by $\tilde{f}=0$, and using form-invariance (this is analogous to the $f=0$ example discussed in section 5.4).  For all such solutions it is then possible to check whether or not the complete positivity condition (\ref{cp}) is satisfied by the corresponding qubit evolution.  

The further reduction of the master equation problem to a first-order differential equation (\ref{gauge}) in section 4.2 provides an interesting subject for further study, as it implies that one may also formulate the general time-dependent oscillator problem in terms of this equation. For example, it can be shown that the particular solution found in section 4.3 corresponds to solving the (inverted) oscillator equation (\ref{osc2}) for  $k(\tau) = -(1-\tau/K)^{-2}$ (for times $\tau < K)$.    More generally, a solution of the `zero-frequency picture' equation (\ref{gauge}) leads to an implicit equation for $k(\tau)$ which must be solved subject to the condition $k(0)=1$.  

Finally, note for the general case of an {\it arbitary} decoherence matrix $\gamma$ and Hamiltonian $H$ in equation (\ref{master}), the question of complete positivity is invariant under unitary transformations, and hence under arbitrary rotations of $\gamma$ and translations of $H$ (see section 4.1).  This implies that the necessary and sufficient conditions for complete positivity, when formulated in terms of the master equation parameters, must be expressible purely in terms of functionals of the three eigenvalues of $\gamma$, with no dependence on  $H$.   Now, for the case of a block-diagonal decoherence matrix, i.e., of form (\ref{general}) with $r=s=0$, these eigenvalues are given by
\[ \lambda_{\pm} = \frac{1}{2}(\gamma_1+\gamma_2) \pm \frac{1}{2} \left[(\gamma_1-\gamma_2)^2 +4f^2 +4g^2\right]^{1/2},~~~~\lambda_3=\gamma_3  . \]
However, the sufficient condition in equation (\ref{bound}) can only be written in terms of these eigenvalues for the case $g=0$. It follows that  a {\it stronger} sufficient condition must exist in general.  It is conjectured, comparing the above eigenvalues with the forms of equations (\ref{a}) and (\ref{w3}), that such a condition may be obtained by replacing $f^2$ by $f^2+g^2$ and $w_3$ by ${w'}$ in equation (\ref{bound}), where $w'$ is defined by the substitution of $\lambda_+ -\lambda_-$ for $g$ in equation (\ref{w3}).

{\bf Acknowledgment}:
I am grateful to Rob Whitney for motivating my interest in `Bloch-diagonal' master equations.

\end{document}